\documentstyle[12pt,epsf]{article}
\renewcommand{\bar}[1]{\overline{#1}}

\begin{document}

\begin{flushright}
GEF-Th-7/2005\\
\end{flushright}

\bigskip\bigskip
\begin{center}
{\large \bf Inequalities between Quark Densities}
\vspace{10pt}
{\bf Elvio Di Salvo\\}
\vspace{5pt}
{Dipartimento di Fisica and I.N.F.N. - Sez. Genova, Via Dodecaneso, 33 \\
 16146 Genova, Italy\\}
\end{center}

\vspace{10pt}
\begin{center} {\large \bf Abstract}
We propose an inequality between the longitudinally polarized density and the transversity 
of a quark in a nucleon. This inequality, whose validity is limited to very small scales, 
is based on considerations about Lorentz transformations and on commonly accepted models. 
Therefore it may be used as a consistency check with other models. It turns out to agree 
with most model predictions. Moreover it allows to establish, thanks to the positivity 
constraint, another inequality between the longitudinally polarized and the unpolarized 
valence quark density. We show that this latter inequality may be extended to any $Q^2$, 
consistently with commonly used factorization schemes and with some nonperturbative 
evolution models. This  inequality finds nontrivial applications to the valence $d$-quark 
densities and is compared with data analyses, with model predictions and with 
parametrizations of quark densities.
\vspace{10pt}
\end{center}

\centerline{PACS numbers: 13.85.Qk, 13.88.+e}

\newpage   
$~~$ Inequalities may be useful in extracting quark densities from data\cite{lss,bt}. In 
particular, important bounds can be fixed by means of the positivity 
constraint\cite{so1,lss}, which gives rise, among other things, to the famous Soffer 
inequality\cite{so}. This may be employed, for example, for determining the behavior of 
the polarized densities for $x \to 1$\cite{bl}. In the present article we propose two 
inequalities concerning polarized and unpolarized densities. The first inequality, which 
regards the transversely and longitudinally polarized densities, may be taken into account 
only at small $Q^2$; it relies on considerations about Lorentz transformations and on a 
property shared by commonly accepted models. This inequality may be used as a consistency 
check with other models and results to be fulfilled by most current model predictions. The 
second inequality is inherent to unpolarized and longitudinally polarized densities. It is 
a consequence of the previous one and of the Soffer inequality and, according to some 
evolution pictures, it may be extended to any $Q^2$. This latter inequality finds 
nontrivial applications to the valence {\it d}-quark densities: it results in accord with 
model predictions, with available data analyses and with some best fits.

We introduce the transverse momentum (tm) longitudinally polarized density $\delta 
q(x,{\bf p}^2_{\perp})$ and the tm transversity $\delta_T q(x,{\bf p_{\perp}})$, whose 
integrals over ${\bf p_{\perp}}$ are, respectively, the functions $\Delta q(x)$ and 
$h_1(x)$ (this latter often denoted as $\Delta_T q(x)$). 

Let us consider $Q^2$ values such that $Q^2 << M^2$, where $M$ is the nucleon rest mass. 
Then a quark can be viewed as a 
constituent quark\cite{ss}. In the rest frame of a polarized nucleon, denote with 
$q_0^{\pm}({\bf p}')$ the probability density for a quark of momentum ${\bf p}'$ $\equiv$ 
$({\bf p_{\perp}}, p'_3)$ to have spin parallel (+) or antiparallel (-) to the nucleon 
spin vector ${\bf S}_0$. A boost parallel to the nucleon spin, and such that the nucleon 
momentum becomes infinite, produces a spin dilution in the quark polarization density, 
which, in the $Q^2$-range considered, turns out to coincide with $\delta q$, {\it i. e.},
\begin{equation}
\delta q(x, {\bf p}^2_{\perp}) = \left[q_0^+({\bf p}') - q_0^-({\bf 
p}')\right] cos\theta_{M}. \label{g1l} 
\end{equation}
Here $x = (p'_0+p'_3)/M$ and $p'_0 = \sqrt{m^2+{\bf p}^{'2}}$. Moreover $\theta_{M}$ 
is the Melosh-Wigner rotation angle\cite{ss,ma}, {\it i. e.},
\begin{equation}
\theta_{M} = arccos\left[\frac{X^2-{\bf p}_{\perp}^2}
{X^2+{\bf p}_{\perp}^2}\right] \label{long}
\end{equation}
and $X = m+p'_0+p'_3$.  
Under the same conditions, a boost from the nucleon rest system, analogous to the previous 
one, but in a direction perpendicular to its spin, produces a less drastic spin dilution. 
Really, in this case the density results in $\delta_T q$, the Melosh-Wigner rotation 
giving\cite{ss}
\begin{equation}
\delta_T q(x, {\bf p}_{\perp}) = \left[q_0^+({\bf p}') - q_0^-({\bf 
p}')\right] D_{\perp}(\theta_{M},\phi). \label{tras}
\end{equation}
Here
\begin{equation}
D_{\perp}(\theta_{M},\phi) = cos^2\frac{\theta_{M}}{2} + 
sin^2\frac{\theta_{M}}{2}(2sin^2\phi-1) \label{tras1}
\end{equation}
and $\phi$ is the azimuthal angle of ${\bf p_{\perp}}$ with respect to the plane 
perpendicular to ${\bf S}_0$. 
Eqs. (\ref{g1l}) and (\ref{tras1}) imply
\begin{equation}
\delta_T q(x, {\bf p}_{\perp}) = \delta q(x, {\bf p}_{\perp}^2)
\frac{D_{\perp}(\theta_{M},\phi)}{cos\theta_{M}}. \label{relaz}
\end{equation}
This, in turn, implies, for $Q^2 << M^2$, the inequality
\begin{equation}
\frac{\delta_T q(x, {\bf p}_{\perp})}{\delta q(x, {\bf p}_{\perp}^2)}\geq 1, \label{ineq1}
\end{equation}
which reduces to equality for a nonrelativistic bound state. Now we make the following 
assumption: 

{\it The difference $q_0^+({\bf p}') - q_0^-({\bf p}')$ is either always positive or 
always negative, except, at most, in a small neighborhood of $x = 1$, where $q_0^-({\bf 
p}') \to 0$}.

This assumption agrees with the predictions of some models\cite{ly,is,ck,dis5}, which 
explain various data concerning the nucleon phenomenology. Moreover the it implies that 
the valence quark density, $\Delta q_{v}$, related to $q_0^+({\bf p}') - q_0^-({\bf p}')$ 
through eq. (\ref{g1l}) and through integration over ${\bf p}_{\perp}$, is either positive 
(for $u$-quarks) or negative (for $d$-quarks) for any $x$ (except, possibly, for $x \simeq 
1$\cite{bl}). As we shall see below, this property may be assumed also at  large $Q^2$, in 
accord with best fits to high energy data\cite{lss}.  

Our assumption implies, together with eq. (\ref{ineq1}),
\begin{equation}
\frac{h_{1}(x)}{\Delta q(x)}\geq 1 \label{ineq1p}
\end{equation}
for $Q^2 << M^2$. Inequality  (\ref{ineq1p}) - which again reduces to an equality for a 
nonrelativistic bound state - is in accord with almost all previous model calculations, 
based on the constituent quark model\cite{ma,ss}, on the bag model\cite{jj,sv}, on light 
cone models\cite{sw} or on the chiral quark model\cite{mg} (see also ref. \cite{bdr} for a 
review). It disagrees only with the calculation based on the chiral quark soliton 
model\cite{dp}. It would be interesting to study the origin of this discrepancy.

At increasing $Q^2$, $h_{1}$ decreases much more rapidly than $\Delta q$, owing to a 
different evolution kernel, therefore the inequalities (\ref{ineq1}) and (\ref{ineq1p}) no 
longer hold true. However, relation (\ref{ineq1p}) gives rise, together with the Soffer 
inequality\cite{so}, to a bound which may be assumed for any $Q^2$, as we shall see in a 
moment. The Soffer inequality reads
\begin{equation}
2|h_{1}(x)| \leq q(x) + \Delta q(x), \label{ineq2}
\end{equation}
where $q(x)$ is the unpolarized quark density. We get from (\ref{ineq1p}) and 
(\ref{ineq2}) 
\begin{equation}
2|\Delta q(x)| \leq q(x) + \Delta q(x) ~~~~~ \ ~~~ (Q^2 << M^2), \label{ineq22}
\end{equation}
which is nontrivial for negative values of $\Delta q(x)$:
\begin{equation}
-3 \Delta q(x) \leq q(x) ~~~~~ \ ~~~ (Q^2 << M^2). \label{ineq3}
\end{equation}
Since $q(x)$ is nonnegative, ineq. (\ref{ineq3}) holds true for {\it any} value of $x$. 
This inequality is especially interesting as regards the {\it valence} $d$-quark in the 
nucleon, whose polarized density is found to be negative for almost all $x$. In 
particular, the nonrelativistic $SU(6)$-symmetric quark model predicts 
\begin{equation}
\Delta d_{v}(x) = -\frac{1}{3} d_{v}(x),  \label{eq3}
\end{equation}
which saturates ineq. (\ref{ineq3}).

Now we show that this inequality, if referred to {\it valence} quarks, {\it i. e.}
\begin{equation}
-3 \Delta q_{v}(x) \leq q_{v}(x), \label{ineq44}
\end{equation}
may be extended to any $Q^2$, consistent with commonly used factorization schemes and 
nonperturbative evolution models. This amounts to showing that the function
\begin{equation}
\phi(x,t) = q_{v}(x,t)+3\Delta q_{v}(x,t), ~~ \ t = ln (Q^2/Q^2_0),
\end{equation}
with $Q^2_0 << M^2$, is positive for any $x$ and $t \geq 0$, given that $\phi(x,0) > 0$, 
as follows from (\ref{ineq3}). To this end, consider evolution equations for $q^{\pm}_v 
(x,t)$, where
\begin{equation}
q^{\pm}_v (x,t) = \frac{1}{2}[q_{v}(x,t) \pm \Delta q_{v}(x,t)]. 
\end{equation}
Taking into account parity conservation, set
\begin{equation}
\frac{d}{dt}q^{\pm}_v (x,t) = \int_x^1 \frac{dy}{y}\left[P^v_n(x,y,t) q^{\pm}_v (y,t)  +
P^v_f(x,y,t) q^{\mp}_v(y,t)\right]. \label{evpm}
\end{equation}
Here $P^v_{n(f)}(x,y,t)$ is the probability density for a quark with initial fractional 
momentum $y$ to evolve into a quark with fractional momentum $x\leq y$, without (with)
helicity flip. These equations present a strong analogy with the Boltzmann 
equation\cite{bst,blt}. Positivity of the quark densities $q^{\pm}_v (x,t)$ 
demands\cite{bst}, for $x < y$, 
\begin{equation}
P_n^v > 0, ~~~~~~~~~ P_f^v > 0,\label{cond}  
\end{equation}
with, at most, singular diagonal terms at $x = y$. Moreover eqs. (\ref{evpm}) imply
\begin{eqnarray}
\frac{d}{dt}q_{v}(x,t) &=& \int_x^1 \frac{dy}{y}P^v(x,y,t)q_{v}(y,t) , \label{e1}
\\
\frac{d}{dt}\Delta q_{v}(x,t) &=& \int_x^1 \frac{dy}{y} \Delta P^v(x,y,t)\Delta q_{v}(y,t) 
, \label{e2}
\\
\frac{d}{dt}\phi(x,t)~~ &=& \int_x^1 \frac{dy}{y} \left[\Delta P^v(x,y,t)\phi(y,t) 
+ 2 P^v_f(x,y,t) q_{v}(y,t)\right]. \label{e3}
\end{eqnarray}
Here $P^v(x,y,t) = P^v_n(x,y,t)+P^v_f(x,y,t)$ and $\Delta P^v(x,y,t) = 
P^v_n(x,y,t)-P^v_f(x,y,t)$. Then from eq. (\ref{e3}) and from the second condition 
(\ref{cond}) it follows that $\phi$ will be positive for $t > 0$ if, for $x < y$, 
\begin{equation}
\Delta P^v > 0, \label{cond1}  
\end{equation}
{\it i. e.}, $P_n^v > P_f^v$. As a byproduct, condition (\ref{cond1}) implies, together 
with eq. (\ref{e2}), that, if $\Delta q_{v}$ is positive (negative) for any $x$ at $t = 
0$, it is  positive (negative) for any $x$ and $t > 0$. Now we discuss to what extent 
condition (\ref{cond1}) may be supported by commonly accepted evolution pictures, 
examining in detail two different situations. 

For sufficiently large $Q^2$ $(> M^2)$ the evolution of the quarks is governed essentially 
by perturbative massless QCD. Therefore helicity conservation implies $P^v_f = 0$ at 
leading order. At higher orders the evolution is scheme dependent\cite{lr}, as the parton 
densities are not directly observable quantities\cite{bst}. However, $P^v_f$ is still 
vanishing in the ${\bar M}{\bar S}$ chirally conserving scheme\cite{vo,ref} (see also 
\cite{lr} and refs. therein), as well as in the JET scheme\cite{lss2,lss}. Moreover, at 
least in the ${\bar M}{\bar S}$ scheme, it is not completely unrealistic to extend down to 
$Q^2 \simeq 0.34 ~~ GeV^2$ the perturbative evolution picture at next-to-leading order, 
respecting the  positivity conditions (\ref{cond}) (see \cite{grsv0} and refs. therein). 

At decreasing $Q^2$ the perturbative quark-gluon interaction is accompanied, and then 
gradually supplanted, by more complex, nonperturbative mechanisms of 
recombination\cite{lm}. In particular, owing to spontaneous chiral symmetry breaking, 
recombination is supposed to give rise, above all, to Goldstone bosons\cite{lm}, 
especially to pions\cite{mg}. Therefore one may assume the prevalent evolution splitting 
for a quark at small $Q^2$ to be $q \to q' \pi$\cite{bf}. In this regime, especially at 
very small $Q^2$, it is not clear whether a DGLAP-type evolution picture is applicable at 
all, since a bound state emission is involved. However, the picture just described seems 
to account satisfactorily for the evolution of the Gottfried sum from very small to large 
$Q^2$\cite{bf}. Therefore it is worth deducing predictions of such a model about the 
evolution of $\phi$. Now each evolution equation (\ref{e1}) to (\ref{e3}) is replaced by a 
system of two coupled equations, whose unknowns are linear combinations of the {\it up} 
and {\it down} valence quark densities, and whose kernels are $2\times 2$ matrices, 
nondiagonal elements referring to charged pion emission. Conditions (\ref{cond}) are 
replaced by $(P_n^v)_{ij} > 0$ and $(P_f^v)_{ij} > 0$ for all matrix elements; moreover 
helicity conservation at forward - {i. e.}, in the most favored direction of emission, 
according to four-momentum and angular momentum conservation - implies $(P_f^v)_{ij} < 
(P_n^v)_{ij}$,
\footnote{If we take  the model of ref. \cite{bf} strictly, we find $(P_f^v)_{ij} = 0$, 
since both the quark and the pion are assumed to be massless.}
which corresponds to condition (\ref{cond1}). Our conclusion is not modified by taking 
into account also a splitting of the type $q \to q' \rho$, which, although contributing to 
helicity flip, is much more unlikely than the one involving the pion\cite{fs}. Nor is our 
argument changed by the assumption that the Gottfried rule fails - due to fluctuations of 
the nucleon into a pion and a baryon\cite{pb} - even at very small scales, as results from 
chiral quark soliton model calculations\cite{qs}. Lastly, due to the linearity of the 
evolution equations, positivity is conserved also in the $Q^2$ range where both 
perturbative and nonperturbative evolution mechanisms coexist.

This completes the argument in favor of ineq. (\ref{ineq44}), which constitutes a 
nontrivial bound. For example, it implies, together with the positivity constraint, $-1/3 
q_{v}(x) \leq \Delta q_{v}(x) \leq q_{v}(x)$, which is stronger than the inequality 
$|\Delta q(x)| \leq q(x)$, usually taken into account in the fits to data of polarized 
deep inelastic scattering\cite{lss}.   

Ineq. (\ref{ineq44}) agrees with the predictions of some models, like the constituent 
quark model\cite{is} and the Carlitz-Kaur model\cite{ck} (see also ref. \cite{qr}). 
Moreover, as regards $d$-quarks, integrating over $x$ from 0 to 1 both sides of that 
inequality yields 
\begin{equation}
\Delta D_v \geq -1/3, \label{ineq4}
\end{equation} 
where $\Delta D_v$ is the first moment of $\Delta d_{v}(x)$. Bound (\ref{ineq4}) is 
fulfilled by a lattice calculation\cite{goe}; it is also in agreement with 
HERMES\cite{her} and SMC\cite{smc} data analyses, where one has assumed a flavor symmetric 
sea polarization. Concerning best fits, this bound agrees with the one by Leader, Sidorov 
and Stamenov\cite{lss}, who assume an $SU(3)$-symmetric polarized sea. The accord is even 
improved, if an asymmetry is introduced in the  polarized sea, either by means of an {\it 
ad hoc} parameter\cite{lss3,tb,bt1}, or as a consequence of a Pauli-blocking 
ansatz\cite{grsv}.
On the contrary, bound (\ref{ineq4}) is not respected by the fit in ref. \cite{bt}, where 
no constraints are assumed for the sea. Of course, as pointed out in refs. \cite{tb,bt1}, 
the splitting of a polarized quark density into valence and sea contributions is strongly 
model dependent. But ineq. (\ref{ineq44}) may reduce such a dependence, consistently with 
commonly used factorization schemes - adherent, as far as possible, to a parton model 
description\cite{cfp,vo} - and with nonperturbative models at $Q^2 << M^2$. 

To summarize, we have proposed two inequalities, (\ref{ineq1p}) and (\ref{ineq44}). The 
former inequality, which relies on general considerations about Lorentz transformations 
and on a property shared by commonly accepted models, is found to agree with almost all 
model predictions about $h_1(x)$ and $\Delta q(x)$ at very small scales. The latter 
inequality, deduced from the former one, from Soffer's inequality and successful evolution 
pictures, agrees with model predictions, with analyses of available data and with some 
best fits.      

\vskip 0.30in

\centerline{\bf Acknowledgements}
The author is grateful to his friend G. Ridolfi for fruitful discussions.


\vskip 0.30in


\begin{thebibliography}{99}

\bibitem{lss} E. Leader, A.V. Sidorov and D.B. Stamenov, Eur. Phys. J. C {\bf 23} (2002) 
479; Phys. Lett. B {\bf 445} (1998) 232

\bibitem{bt} J. Bartelski and S. Tatur, Phys. Rev. D {\bf 65} (2002) 034002; 
hep-ph/0205089

\bibitem{so1} J. Soffer, Phys. Rev. Lett. {\bf 91} (2003) 092005

\bibitem{so} J. Soffer, Phys. Rev. Lett. {\bf 74} (1995) 1292

\bibitem{bl} M. Boglione and E. Leader, Phys. Rev. D {\bf 61} (2000) 114001 

\bibitem{ss} I. Schmidt and J. Soffer,  Phys. Lett. B {\bf 407} (1997) 331

\bibitem{ma} B.-Q. Ma, J. Phys. G: Nuc. Part. Phys. {\bf 17} (1991) L53

\bibitem{ly} A. Le Yaouanc, L. Oliver, O. P\`ene and J.C. Reynal, Phys. Rev.{\bf 9} (1974) 
2636; {\bf 12} (1975) 2137

\bibitem{is} N. Isgur, Phys. Rev. D {\bf 59} (1999) 034013

\bibitem{ck} R. Carlitz and J. Kaur, Phys. Rev. Lett. {\bf 38} (1977) 673

\bibitem{dis5} E. Di Salvo, J. Phys. G: Nuclear and Particle Physics {\bf 16} L285 (1990)

\bibitem{jj} R.L. Jaffe and X. Ji, Phys. Rev. Lett. {\bf 67} (1991) 552;
Nucl. Phys. B {\bf 375} (1992) 527

\bibitem{sv} S. Scopetta and V. Vento, Phys. Lett. B {\bf 424} (1997) 31

\bibitem{sw} K. Suzuki and W. Weise, Nucl. Phys. A {\bf 634} (1998) 141
 
\bibitem{mg} A. Manohar and H. Georgi, Nucl. Phys. B {\bf 234} (1984) 189

\bibitem{bdr} V. Barone, A. Drago and P. Ratcliffe, Phys. Rept. {\bf 359} (2002) 1

\bibitem{dp} D.I. Diakonov and V.Yu. Petrov, Nucl. Phys. B {\bf 272} (1986) 457; D.I. 
Diakonov, V.Yu. Petrov and P.V. Pobylitsa, Nucl. Phys. B {\bf 306} (1988) 809

\bibitem{bst} C. Bourrely, J. Soffer and O. Teryaev, Phys. Lett. B {\bf 420} (1998) 375

\bibitem{blt} C. Bourrely, E. Leader and O. Teryaev, hep-ph/9803238, Talk given at the VII 
Workshop on High energy Spin Physics (SPIN-97), Dubna, July 7-12, 1997

\bibitem{lr} B. Lampe and E. Reya: Phys. Rep. {\bf 332} (2000) 1

\bibitem{vo} W. Vogelsang, Phys. Rev. D {\bf 54} (1996) 2023; Nucl. Phys. B {\bf 475} 
(1996) 47

\bibitem{ref} See refs. \cite{lr,vo} and refs. therein

\bibitem{lss2} E. Leader, A.V. Sidorov and D.B. Stamenov, Phys. Rev. D {\bf 58} (1998) 
14028;
see also refs. therein

\bibitem{grsv0} M. Glueck, E. Reya, M. Stratmann and W. Vogelsang, Phys. Rev. D {\bf 53}
(1996) 4775

\bibitem{lm} M. Lavelle and D. McMullan, Phys. Rep. {\bf 279} (1997) 1

\bibitem{bf} R. Ball  and  S. Forte, Nucl. Phys. B {\bf 425} (1994) 516

\bibitem{fs} R.J. Fries and A. Shaefer, Phys. Lett. B {\bf 443} (1998)

\bibitem{pb} see, {\it e. g.}, ref. \cite{fs} and refs. therein

\bibitem{qs} D. Diakonov {\it et al.}, Nucl. Phys. B {\bf 480} (1996) 341

\bibitem{qr} J. Qiu, G.P. Ramsey, D. Richards and D. Sivers, Phys. Rev. D {\bf 41} (1990) 
65 

\bibitem{goe} M. Goeckeler et al., Phys. Lett. B {\bf 414} (1997) 340

\bibitem{her} HERMES coll., K. Ackerstaff et al., Phys. Lett. B {\bf 464} (1999) 123

\bibitem{smc} SMC coll., B. Adeva et al., Phys. Lett. B {\bf 420} (1997) 180

\bibitem{lss3} E. Leader, A.V. Sidorov and D.B. Stamenov, Phys. Rev. D {\bf 58} (1998) 
14028 

\bibitem{tb} S. Tatur, J. Bartelski and M. Kurzela, Acta Phys. Pol. {\bf 31} (2000) 647

\bibitem{bt1} J. Bartelski and S. Tatur, Acta Phys. Pol. {\bf 32} (2001) 2101

\bibitem{grsv} M. Glueck, E. Reya, M. Stratmann and W. Vogelsang, Phys. Rev. D {\bf 63} 
(2001) 094005

\bibitem{cfp} C. Curci, W. Furmanski and R. Petronzio, Nucl. Phys. B {\bf 175} (1980) 27 

\end{thebibliography}
\end{document}